\documentclass[%
 reprint,
 amsmath,amssymb,
 aps,
prd,
floatfix,
nofootinbib
]{revtex4-1}

\usepackage{graphicx}
\usepackage{dcolumn}
\usepackage{bm}
\usepackage{amsmath,amsfonts,amssymb,slashed,upgreek,xcolor}
\usepackage{multirow}
\usepackage{enumitem}
\usepackage{cancel}
\usepackage{hyperref}
\usepackage{braket}
\usepackage{lipsum}
\usepackage{float}

\newcommand{\HH}{\mathcal{H}}
\newcommand{\hJ}{\hat{J}}
\newcommand{\hb}{\hat{b}}

\newcommand{\bk}{\mathbf{k}}
\newcommand{\bn}{\mathbf{n}}
\newcommand{\bV}{\mathbf{V}}

\newcommand{\itt}[1]{\textcolor{black}{#1}}

\newcolumntype{C}[1]{>{\centering\arraybackslash}p{#1}}

\begin{document}

\title{Combining gravitational lensing and gravitational redshift to measure the anisotropic stress with future galaxy surveys}

\author{Isaac Tutusaus}
\email{isaac.tutusaus@irap.omp.eu}
\affiliation{Institut de Recherche en Astrophysique et Plan\'etologie (IRAP), Universit\'e de Toulouse, CNRS, UPS, CNES, 14 Av. Edouard Belin, F-31400 Toulouse, France}
\affiliation{D\'epartment de Physique Th\'eorique and Center for Astroparticle Physics, Universit\'e de Gen\`eve, Quai E. Ansermet 24, CH-1211 Gen\`eve 4, Switzerland}

\author{Daniel Sobral Blanco}
\email{daniel.sobralblanco@unige.ch}
\affiliation{D\'epartment de Physique Th\'eorique and Center for Astroparticle Physics, Universit\'e de Gen\`eve, Quai E. Ansermet 24, CH-1211 Gen\`eve 4, Switzerland}

\author{Camille Bonvin}
\email{camille.bonvin@unige.ch}
\affiliation{D\'epartment de Physique Th\'eorique and Center for Astroparticle Physics, Universit\'e de Gen\`eve, Quai E. Ansermet 24, CH-1211 Gen\`eve 4, Switzerland}

\date{\today}

\begin{abstract}
Galaxy surveys provide one of the best ways to constrain the theory of gravity at cosmological scales. They can be used to constrain the two gravitational potentials encoding time, $\Psi$, and spatial, $\Phi$, distortions, which are exactly equal at late time within general relativity. Hence, any small variation leading to a nonzero anisotropic stress, i.e.\ a difference between these potentials, would be an indication for modified gravity. Current analyses usually consider gravitational lensing and redshift-space distortions to constrain the anisotropic stress, but these rely on certain assumptions like the validity of the weak equivalence principle, and a specific time evolution of the functions encoding deviations from general relativity. In this work, we propose a reparametrization of the gravitational lensing observable, together with the use of the relativistic dipole of the correlation function of galaxies to directly measure the anisotropic stress with a minimum amount of assumptions. We consider the future Legacy Survey of Space and Time of the Vera C. \itt{Rubin} Observatory and the future Square Kilometer Array, and show that combining gravitational lensing and gravitational redshift with the proposed approach we will achieve model-independent constraints on the anisotropic stress at the level of $\sim 20\,\%$. 
\end{abstract}

\pacs{Valid PACS appear here}
\maketitle

\section{Introduction} 
\label{sec:intro}

Testing the laws of gravity at cosmological scales is one of the main goals of current and future large-scale structure surveys. In particular, one key test of gravity consists in comparing the two gravitational potentials encoding distortions in the geometry of our Universe~\footnote{We use the perturbed Friedmann metric: $\mathrm ds^2 = a^2[-(1+2 \Psi)\mathrm d \tau^2 + (1- 2 \Phi)\mathrm d \mathbf{x}^2]$, where $\tau$ denotes conformal time.}: namely the time distortion, $\Psi$, and the spatial distortion, $\Phi$. In the $\Lambda$CDM model, these two potentials are equal at late times~\footnote{Note that massive neutrinos generate a difference between these two potentials at late time, but the amplitude of this difference has been shown to be very small: $(\Phi+\Psi)/(2\Psi)-1\simeq 10^{-5}$~\cite{Adamek:2017uiq}.}. Most dark energy models do preserve this equality, whereas theories of gravity beyond general relativity (GR) generically induce a difference between $\Psi$ and $\Phi$, called anisotropic stress~\cite{Motta:2013cwa,Amendola:2016saw}. Measuring the anisotropic stress provides therefore a stringent way of testing dark energy and modified gravity theories~\cite{Song:2010fg,EuclidTheoryWorkingGroup:2012gxx,Amendola:2016saw}. A detection of a nonzero anisotropic stress is often referred to as a \emph{smoking gun} for modified gravity. In particular, it would allow us to distinguish between interactions in the dark sector (that preserve the equality between $\Phi$ and $\Psi$) and a modification of gravity.

At present, two methods have been used to measure the anisotropic stress.
The first one consists in combining measurements of galaxy peculiar velocities obtained from redshift-space distortions (RSD), with measurements of the Weyl potential, $(\Phi+\Psi)/2$, obtained from gravitational lensing. Assuming that galaxies obey Euler's equation, RSD measurements can be translated into a measurement of the time distortion, $\Psi$. Comparing this with the Weyl potential allows one to measure the anisotropic stress, which, at the precision of current surveys, is consistent with zero~\cite{DES:2018ufa,eBOSS:2020yzd,DES:2022ygi}. This method is very neat and powerful but it clearly fails if galaxies do not obey Euler's equation, which is the case if dark matter is affected by a fifth force (like for example in coupled quintessence models~\cite{Amendola:1999er}). As shown in~\cite{levon}, this breaking of the equivalence principle invalidates the method. It would indeed lead to a measurement of an effective non-zero anisotropic stress from RSD and lensing, even if $\Phi=\Psi$.

The second method to constrain the anisotropic stress consists in looking at the speed of propagation of gravitational waves (GWs). In scalar-tensor theories of gravity (Horndeski theories~\cite{Horndeski:1974wa}), one of the parameters that generates a non-zero anisotropic stress also governs the speed of propagation of GWs. Combining the constraints on this speed obtained through the GW and electromagnetic observations of the binary neutron stars system GW170817~\cite{LIGOScientific:2017vwq}, with the consideration that dark energy perturbations in Horndeski may become unstable in the presence of GWs~\cite{Creminelli:2019kjy} (which may further limit the parameter space), leads to constraints on the anisotropic stress of the order of a few percent~\cite{Noller:2020afd}. This elegant method, which combines GWs propagation with large-scale structure, is however applicable only to a specific class of modified gravity theories and is therefore not model-independent. As before, it relies on the validity of the weak equivalence principle in Horndeski theories. Moreover, it assumes that the speed of GWs is independent of frequency and of time, something that has not been validated observationally.

In this paper, we propose an alternative method to measure the anisotropic stress, which has the key advantage of being model-independent. Our method combines direct measurements of the time distortion, $\Psi$, which can be detected via the effect of gravitational redshift, with measurements of the Weyl potential from gravitational lensing, to \emph{directly} probe the relation between $\Psi$ and $\Phi$. This method does not rely on a specific model of gravity, nor on a particular behavior of dark matter: it uses the data in a completely agnostic way to extract measurable quantities and compare them. In particular, contrary to measurements from RSD and weak lensing, this method is also valid if the weak equivalence principle is violated, i.e.\ if dark matter obeys a fifth force. 

\itt{Our analysis follows the method presented in~\cite{Sobral-Blanco:2021cks}, where it was shown that combining multipoles of the galaxy power spectrum with the galaxy-galaxy lensing power spectrum provides a direct model-independent determination of $\eta\equiv \Phi/\Psi$. Here we apply this method to the coming generation of surveys to determine the precision with which $\eta$ can be measured. Note that we use here the correlation function and the angular power spectrum, instead of the power spectrum, to properly account for wide-angle effects~\cite{Bonvin:2013ogt,Tansella:2017rpi}.}
Our forecasts show that combining measurements of gravitational redshift from the Square Kilometer Array with measurements of the Weyl potential from \itt{the Legacy Survey of Space and Time of the Vera C. Rubin Observatory (LSST)} will allow us to constrain the anisotropic stress, through $2/(1+\eta)=2/(1+\Phi/\Psi)$, with a precision of $\sim 20\,\%$. 

The rest of the paper is organized as follow: in Sec.~\ref{sec:modeling} we describe the modeling considered for both the gravitational lensing and gravitational redshift observables, as well as how they are combined to constrain the anisotropic stress. In Sec.~\ref{sec:methodology} we present the specifications of the future galaxy surveys considered and the methodology used to forecast the constraints on the different parameters under study. We then present the main results of the analysis in Sec.~\ref{sec:results} for both observables and their combination, and we conclude in Sec.~\ref{sec:conclusions}.

\section{Modeling} 
\label{sec:modeling}
We start by deriving the modeling used to express the predictions for our observables in a model-independent way, and thus directly test the anisotropic stress with a minimum amount of assumptions. We first consider the gravitational lensing observable related to the Weyl potential; then the gravitational redshift related to $\Psi$; and finally the ratio related to the anisotropic stress. 

\subsection{Gravitational lensing: \texorpdfstring{$\Phi+\Psi$}{Lg}}
\label{sec:sum}

The standard method to model the evolution of the two gravitational potentials consists in relating them to the matter density, allowing for a non-zero anisotropic stress and a modification to Poisson's equation. Here we develop a different method, which, as we will show, allows us to measure directly the evolution of $\Phi+\Psi$ with redshift.

We start by reviewing how $\Phi+\Psi$ evolves with redshift in $\Lambda$CDM. We define the transfer function of any field, $F$, as
\begin{align}
    F(\bk, z)=T_{F}(k,z)\Psi_{\rm in}(\bk)\, ,
\end{align}
where $\Psi_{\rm in}$ denotes the primordial gravitational potential generated by inflation. In $\Lambda$CDM the transfer function of $\Phi+\Psi$ can be related to that of the matter density through
\begin{equation}\label{eq1}
    T_{\Phi+\Psi}(k,z)=2T_{\Phi}(k,z)=-3\Omega_{\rm m}(z)\left[\frac{\mathcal{H}(z)}{k}\right]^2T_{\delta}(k,z)\,,
\end{equation}
where $\Omega_{\rm m}(z)$ stands for the matter density of the Universe at a given redshift $z$, $\mathcal{H}$ denotes the Hubble rate in conformal time, and $\delta$ is the matter density fluctuations $\delta=\delta\rho/\bar \rho$. The matter density transfer function is usually split into a linear part and a boost factor, $B$, which accounts for the nonlinear evolution of matter fluctuations at small scales
\begin{equation}
    T_{\delta}(k,z)=T_{\delta}^{\rm lin}(k,z)\sqrt{B(k,z)}\,.\label{eq:boost}
\end{equation}

In $\Lambda$CDM, the growth of matter density fluctuations in the linear regime is scale-independent at late times (once radiation is negligible)~\footnote{The only scale-dependence within \itt{general relativity (GR)} is indeed due to massive neutrinos and it is very small (see, e.g., \cite{ISTFpaper}).}, and we can therefore relate the transfer function at redshift $z$ to a redshift $z_*$ through the growth function $D_1(z)$
\begin{equation}
    T_{\delta}^{\rm lin}(k,z)=\frac{D_1(z)}{D_1(z_*)}T_{\delta}^{\rm lin}(k,z_*)\,.\label{eq:growth}
\end{equation}
Inserting Eqs.~\eqref{eq:boost} and~\eqref{eq:growth} into~\eqref{eq1} we find
\begin{align}\label{eq5}
    T_{\Phi+\Psi}(k,z)=&\frac{\mathcal{H}^2(z)\Omega_{\rm m}(z)D_1(z)}{\mathcal{H}^2(z_*)\Omega_{\rm m}(z_*)D_1(z_*)}\\
    &\times\sqrt{B(k,z)}T_{\Phi+\Psi}(k,z_*)\,.\nonumber
\end{align}
Equation~\eqref{eq5} tells us that, in $\Lambda$CDM, the evolution of $\Phi+\Psi$ is governed by $\Omega_{\rm m}(z)D_1(z)$, i.e.\ that $\Phi+\Psi$ follows directly the evolution of the density. 

In the case of modified gravity, however, the two equalities present in Eq.\,(\ref{eq1}) are generically modified: first $\Phi$ and $\Psi$ can be different from each other, and second $\Phi$ may not be related to the density, $\delta$, via Poisson's equation. Therefore, generically, the evolution of $\Phi+\Psi$ will differ from the evolution of the density, governed by $D_1$. To account for this, without restricting ourselves to any particular model of gravity, we replace $\Omega_{\rm m}(z)D_1(z)$ in Eq.\,(\ref{eq5}) by an agnostic function $J(k,z)$. This function encodes the growth of $\Phi+\Psi$, and by treating it as a \emph{new degree of freedom}, that can be directly measured from the data and that is independent from the growth function $D_1$, we effectively allow for any deviations from GR.

We define the redshift $z_*$ to be well in the matter era, before the accelerated expansion of the Universe started. We assume that, at that redshift, GR is recovered. In other words, we expect structures to grow as in GR, when the background evolution of the Universe behaves as in GR. We therefore set $J(k,z_*)=\Omega_{\rm m}(z_*)D_1(z_*)=D_1(z_*)$, since $\Omega_{\rm m}(z_*)=1$ in the matter era. With this, the transfer function for $\Phi+\Psi$ evolves as
\begin{align}\label{eq:evolJ}
    T_{\Phi+\Psi}(k,z)=&\frac{\mathcal{H}^2(z)J(k,z)}{\mathcal{H}^2(z_*)D_1(z_*)}\sqrt{B(k,z)}T_{\Phi+\Psi}(k,z_*)\,. 
\end{align}

Combining Eq.~\eqref{eq:evolJ} with Eq.~\eqref{eq:boost} and~\eqref{eq:growth}, and using that at $z_*$ GR is recovered, i.e.
\begin{equation}
    T_{\delta}^{\rm lin}(k,z_*)=-\frac{1}{3}\left[\frac{k}{\mathcal{H}(z_{*})}\right]^2T_{\Phi+\Psi}(k,z_*)\,, 
\end{equation}
we can relate the transfer function of $\Phi+\Psi$ to the density transfer function at $z_*$ via
\begin{align}
    T_{\Phi+\Psi}(k,z)&=-3\left[\frac{\mathcal{H}(z)}{k}\right]^2J(k,z)\sqrt{B(k,z)}\frac{T_{\delta}^{\rm lin}(k,z_*)}{D_1(z_*)}\label{eq:Tweyl}\, .
\end{align}

Let us now determine how the function $J$ can be measured with gravitational lensing. Gravitational lensing can be measured either through shear-shear correlations or through density-shear correlations (the so-called galaxy-galaxy lensing). In this paper we focus on the latter, since, as we will see, it allows us to probe directly the function $J$ at the redshift of the lenses. In addition, we include in our analysis the galaxy-galaxy correlations since it allows us to break the degeneracy between the galaxy bias and the function $J$.

The harmonic power spectra of the galaxy-galaxy lensing can be computed as
\begin{align}
    C_{\ell}&^{\!\!\Delta\kappa}(z_i,z_j)=\int \text{d}z\,n_i(z)b_i(z)\int\text{d}z'n_j(z')C_{\ell}^{\delta\kappa}(z,z')\nonumber\\
    =&-\frac{A}{\pi}\int\text{d}z\,n_i(z)b_i(z)\int\text{d}z'n_j(z')\int\frac{\text{d}k}{k}\left(\frac{k}{k_*}\right)^{n_{\rm s}-1}\nonumber\\
    &\times T_{\delta}(k,z)j_{\ell}(k\chi)\frac{1}{\chi'}\int_0^{\chi'}\text{d}\chi''\frac{\chi'-\chi''}{\chi''}\ell(\ell+1)\nonumber\\
    &\times T_{\Phi+\Psi}(k,\chi'')j_{\ell}(k\chi'')\,. \label{eq:galaxylensing}
\end{align}
Here $\Delta$ denotes the galaxy over-density, evaluated at the effective redshift of the tomographic bin $i$
\begin{align}
\Delta(z_i,\bn_i)=b_i\delta(z_i,\bn_i)+\frac{1}{\HH}\partial_\chi(\bV\cdot\bn_i)\, , \label{eq:Delta}    
\end{align}
where $\bn_i$ is the direction of the pixel $i$, $\bV$ is the galaxy peculiar velocity, $b_i$ is the linear bias, and $\partial_\chi$ denotes a derivative with respect to the comoving distance $\chi$. $\kappa$ represents the convergence, evaluated in pixel $j$ with effective redshift $z_j$. It is related to the Weyl potential by
\begin{align}
\kappa(z_j,\bn_j) & =  \int_0^{\chi_j}\!d \chi\,\frac{\chi_j-\chi}{2 \chi_j\chi}\,\Delta_\Omega(\Phi+\Psi)\big(\chi,\bn_j\big)\, ,  \label{eq:kappa}
\end{align}
where $\chi_j\equiv \chi(z_j)$ and $\Delta_\Omega$ is the Laplace operator on the sphere. In Eq.~\eqref{eq:galaxylensing}, $n_i$ and $n_j$ denote the galaxy distribution function of the lenses and sources, respectively, and $j_{\ell}$ stands for the spherical Bessel function of order $\ell$. The parameters $A$, $n_{\rm s}$, and $k_*$ denote the amplitude, spectral index, and pivot scale of the primordial power spectrum defined through\footnote{Note that $A$ is related to the amplitude $A_{\rm s}$ defined in Planck through $A=8\pi^2 A_{\rm s}/9$.} $k^3\langle\Psi_{\rm in}(\bk)\Psi_{\rm in}(\bk')\rangle=(2\pi)^3 A (k/k_*)^{n_{\rm s}-1}\delta(\bk+\bk')$. Note that in Eq.~\eqref{eq:galaxylensing}, we have neglected the correlation between the convergence and RSD [the second term in Eq.~\eqref{eq:Delta}], as is done, e.g.,\ in~\cite{ISTFpaper}, since those are subdominant for thick tomographic bins. 

Using the Limber approximation and inserting Eqs.~\eqref{eq:Tweyl},~\eqref{eq:boost}, and~\eqref{eq:growth} we can express the harmonic power spectra as
\begin{align}\label{eq8}
    C_{\ell}^{\Delta\kappa}(z_i&,z_j)=\int\text{d}z\,n_i(z)b_i(z)\int\text{d}z'n_j(z')\frac{\chi'-\chi}{\chi\chi'}\nonumber\\
    &\times\frac{3\ell(\ell+1)}{2(\ell+1/2)^2}\mathcal{H}^2(z)\frac{J(z)}{D_1(z)} P_{\delta\delta}\left(k_{\ell},\chi\right)\,,
\end{align}
where $k_{\ell}\equiv (\ell+1/2)/\chi$. Here we have neglected the $k$-dependence of the functions $D_1$ and $J$. This is a common approximation in large-scale structure analyses, see, e.g.~\cite{eBOSS:2020yzd}, which is motivated by the fact that in the quasistatic approximation, many models of modified gravity lead to a scale-independent growth of structure~\cite{Gleyzes:2015pma,Gleyzes:2015rua,Raveri:2021dbu}. However, the methodology can be generalized to include the scale-dependence in case of need. From Eq.~\eqref{eq8}, we see that $J$ is evaluated at the redshift of the lenses, whose distribution is given by $n_i(z)$. 

Writing now the density power spectrum as
\begin{equation}
    P_{\delta\delta}\left(k_{\ell},\chi\right)=\left[\frac{D_1(z)}{D_1(z_*)}\right]^2P_{\delta\delta}^{\rm lin}\left(k_{\ell},\chi_*\right)B\left(k_{\ell},\chi\right)\,,
\end{equation}
we can rewrite Eq.\,(\ref{eq8}) as
\begin{align}
    C_{\ell}^{\Delta\kappa}(z_i,z_j)=&\frac{3}{2}\int\text{d}z\,n_i(z)\mathcal{H}^2(z)\hat{b}_i(z)\hat{J}(z)\nonumber\\
    &\times B\left(k_{\ell},\chi\right)\frac{P_{\delta\delta}^{\rm lin}\left(k_{\ell},\chi_*\right)}{\sigma_8^2(z_*)}\nonumber\\
    &\times \int\text{d}z'n_j(z')\frac{\chi'(z')-\chi(z)}{\chi(z)\chi'(z')}\,, \label{eq:DeltakappaJ}
\end{align}
where we have defined
\begin{equation}\label{eq11}
    \hat{J}(z)\equiv\frac{J(z)\sigma_8(z)}{D_1(z)}=\frac{J(z)\sigma_8(z_*)}{D_1(z_*)}\,,
\end{equation}
and
\begin{equation}\label{eq14}
    \hat{b}_i(z)\equiv b_i(z)\sigma_8(z)\,.
\end{equation}

From Eq.~\eqref{eq:DeltakappaJ}, we see that galaxy-galaxy lensing is affected by four distinct ingredients:
\begin{enumerate}[itemsep=-0.02cm, topsep=0.1cm]
    \item The background expansion of the Universe, through the Hubble function $\mathcal{H}(z)$ and the comoving distance $\chi(z)$.
    \item The density fluctuations at redshift $z_*$, before acceleration started.
    \item The nonlinear boost factor $B$.
    \item The evolution of density and gravitational potential at late times, through the two functions $\hat{b}$ and $\hat{J}$.
\end{enumerate}
Ingredients 1 and 2 are tightly constrained by cosmic microwave background (CMB) measurements. In the following we therefore fix them, for simplicity, using the latest Planck $\Lambda$CDM values for $A_{\rm s}$, $n_{\rm s}$, $\Omega_{\rm m,0}$, $\Omega_{\rm b,0}$, and $h$~\cite{2020A&A...641A...6P}~\footnote{Note that here we assume that the background evolution is consistent with $\Lambda$CDM predictions, since this has been so far confirmed by observations. This assumption is common in large-scale structure analyses, see, e.g.,~\cite{eBOSS:2020yzd}.}. But we note that the methodology described in this work would allow for a combination of galaxy survey data with CMB observations to constrain all parameters together, including the cosmological and the $\hJ$ and $\hat{b}$ parameters. 
Ingredient 3 depends in principle on the theory of gravity. However, the standard method to infer the nonlinear boost is by generating cosmological simulations, which require the choice of a specific modified gravity model. Therefore, it is not possible to obtain a general nonlinear boost factor for our parametrization. In this work, we follow the approach considered in the Dark Enery Survey analysis~\cite{DES:2022ygi} and keep the standard halofit nonlinear boost~\cite{2012ApJ...761..152T}, while limiting the range of scales used in the analysis to avoid entering deeply into the nonlinear regime. We also consider a more stringent scale cut to assess the impact of this choice of boost in our results, showing that it is subdominant and constitutes, therefore, an acceptable approximation. 
Finally, ingredient 4 is what we want to measure in this work. In practice, we assume $\hJ$ and $\hb$ to be free parameters with a constant value within each tomographic bin, and we focus on constraining them.

As can be seen from Eq.\,(\ref{eq:DeltakappaJ}), once we consider $\hJ$ and $\hb$ as free parameters with a constant amplitude in each tomographic bin, their product is fully degenerate. Therefore, as mentioned at the beginning of the section, we do not consider galaxy-galaxy lensing measurements alone, but rather their combination with galaxy clustering, using the same photometrically-selected lenses. This is the so-called 2x2pt analysis. 

The harmonic power spectra for galaxy clustering, using the Limber approximation, is given by
\begin{align}\label{eq:DeltaDeltaJ}
    C_{\ell}^{\Delta\Delta}(z_i,z_j)=&\int\text{d}z\,n_i(z)n_j(z)\itt{\frac{\mathcal{H}(z)(1+z)}{\chi^2(z)}}\hat{b}_i(z)\hat{b}_j(z)\nonumber\\
    &\times B\left(k_{\ell},\chi\right)\frac{P_{\delta\delta}^{\rm lin}\left(k_{\ell},\chi_*\right)}{\sigma_8^2(z_*)}\,.
\end{align}
Combining this expression for the galaxy clustering observable with the galaxy-galaxy lensing observable in Eq.\,(\ref{eq:DeltakappaJ}), we can break the degeneracy between $\hJ$ and $\hb$ and constrain both sets of parameters at the same time.

This procedure allows us to measure the evolution of the gravitational potentials, $\Phi+\Psi$, in a model-independent way. The only assumption underlying such an analysis is that at high redshift $z_*$, before acceleration started, we recover GR. Note that this procedure is similar to the one used in RSD measurements, where the density power spectrum at $z_*$ is constrained by CMB measurements, and RSD are used to measure $\hb$ and $\hat{f}\equiv f\sigma_8$ in a model-independent way in each redshift bin, see, e.g.,~\cite{eBOSS:2020yzd}.

\subsection{Comparison with the standard \texorpdfstring{$\mu-\Sigma$}{Lg} parametrization}
Before moving to the next observable, let us compare our approach with the standard parametrization used in weak lensing analyses. Modifications to GR are usually encoded into two phenomenological functions, $\mu$ and $\eta$, that modify Poisson's equation and the relation between the two gravitational potentials:
\begin{align}
k^2\Psi &=-4 \pi G a^2 \mu(z,k)  \delta\rho\,,  \label{eq:poisson}\\
\Phi&= \eta(z,k) \Psi\,. \label{eq:anis}
\end{align}    
The sum of the gravitational potentials can then be written as
\begin{equation}
k^2(\Phi + \Psi ) =- 3 \HH^2(z)\Omega_{\rm m}(z)\Sigma(z)\delta(k,z)\, ,
\label{eq:sigma}
\end{equation} 
where $\Sigma=\mu(1+\eta)/2$.
With this the galaxy-galaxy lensing power spectra and the galaxy clustering power spectra become
\begin{align}
C_{\ell}^{\Delta\kappa}(z_i,z_j)=&\frac{3}{2}\int\text{d}z\,n_i(z)\mathcal{H}^2(z)\Omega_{\rm m}(z)b_i(z)\Sigma(z)\nonumber\\
    &\times B\left(k_{\ell},\chi\right)P_{\delta\delta}^{\rm lin}\left(k_{\ell},\chi\right)\nonumber\\
    &\times \int\text{d}z'n_j(z')\frac{\chi'(z')-\chi(z)}{\chi(z)\chi'(z')}\,, \label{eq:DeltakappaSigma}\\
    C_{\ell}^{\Delta\Delta}(z_i,z_j)=&\int\text{d}z\,n_i(z)n_j(z)\itt{\frac{\mathcal{H}(z)(1+z)}{\chi^2(z)}}b_i(z)b_j(z)\nonumber\\
    &\times B\left(k_{\ell},\chi\right)P_{\delta\delta}^{\rm lin}\left(k_{\ell},\chi\right)\, .\label{eq:DeltaDeltaSigma}
\end{align}
The galaxy-galaxy lensing power spectra depend directly on the function $\Sigma(z)$. In addition, both the galaxy-galaxy lensing power spectra and the clustering power spectra depend on $\mu$, since the evolution of $\delta$, and consequently the matter power spectrum at redshift $z$, are sensitive to $\mu$.

\begin{table}[t]
\centering
\caption{Comparison of the parameters used in the standard $\mu-\Sigma$ approach and in our approach.}
\label{tab:comp}
\begin{tabular}{C{3.8cm}|C{4.5cm}}
\hline
$\mu-\Sigma$ approach  & Our approach \\ 
\hline
$A_s,\ n_s,\ \Omega_{\rm b, 0},\ \Omega_{\rm m,0},\ h$   & $A_s,\ n_s,\ \Omega_{\rm b, 0},\ \Omega_{\rm m,0},\ h$  \\[2.3pt]
$b(z)$ & $\hb(z)=b(z)\sigma_8(z)$\\[2.3pt]
$\Sigma(z)$ & $\hat{J}(z)=\Sigma(z)\Omega_{\rm m}(z)\sigma_8(z)$\\[2.3pt]
$\mu(z)$ & adding RSD: $\hat{f}(z)=f(z)\sigma_8(z)$
\end{tabular}
\end{table}

Comparing Eq.~\eqref{eq:DeltakappaSigma} with Eq.~\eqref{eq:DeltakappaJ} we see that
\begin{align}
\hat{J}(z)=\Sigma(z)\Omega_{\rm m}(z)\sigma_8(z)\, . \label{eq:SigmaJ}
\end{align}
In Table~\ref{tab:comp}, we list the set of free parameters in the standard $\mu-\Sigma$ parametrization and in our parametrization. The parameters in the first line are the standard cosmological parameters that are best determined by \itt{the} CMB, and that are the same in both cases. In our forecasts, we will keep them fixed for simplicity, but in practice, in both approaches, one can combine lensing with CMB measurements to constrain these parameters. In addition to these standard parameters, the $\mu-\Sigma$ parametrization has three free parameters per redshift bin ($b, \Sigma, \mu$), whereas our parametrization has two free parameters per redshift \itt{bin} ($\hat{b}, \hat{J}$). One could conclude that the $\mu-\Sigma$ parametrization is more powerful since it allows us to measure one more parameter at each redshift. This is however not the case, since $b$, $\Sigma$\itt{,} and $\mu$ are strongly degenerated in Eq.~\eqref{eq:DeltakappaSigma} and~\eqref{eq:DeltaDeltaSigma}. To break the degeneracy it is necessary to add new information, through RSD, which are sensitive to $\mu$ and $b$. In our approach, adding RSD would add one new free function $\hat{f}(z)=f(z)\sigma_8(z)$, that can directly be measured from RSD and leads to the same number of free parameters in both approaches.

From this we see that the first key property of our approach is that it separates clearly the information that can be measured with the 2x2pt lensing measurements, from the information that can be measured from RSD. The 2x2pt data measure the evolution of $\Phi+\Psi$ and the evolution of the galaxy density, whereas RSD measure the evolution of the velocity (and again the evolution of the galaxy density). In the standard $\mu-\Sigma$ parametrization this separation cannot be applied since the parameters $b, \Sigma$\itt{,} and $\mu$ can only be measured by combining lensing with RSD.

The second specificity of our approach is to allow \itt{for} a direct measurement of the parameters $\hat{J}$ and $\hat{b}$ in each redshift bin. In contrast, in the $\mu-\Sigma$ approach, it is not straightforward to measure $\mu(z)$ in each redshift bin. The reason for this is that $\mu$ enters in Eqs.~\eqref{eq:DeltakappaSigma} and~\eqref{eq:DeltaDeltaSigma} through its impact on the matter power spectrum, $P_{\delta\delta}(k,z)$, which depends on the whole time evolution of $\mu(z)$. In other words, to constrain $\mu(z)$ from the 2x2pt data and RSD, we need to solve the evolution equation for the matter density $\delta$, which reads
\begin{align}
&\delta''(k,a)+ \left(1+\frac{\HH'(a)}{\HH(a)}\right) \delta'(k,a)\nonumber\\
&-\frac{3}{2}\frac{\Omega_{m,0}}{a}\left(\frac{\HH_0}{\HH(a)}\right)^2 \mu(a)\,\delta(k,a)=0\,, \label{eq:deltaevol}
\end{align}
where a prime denotes derivatives with respect to $\ln a$. This means that to measure $\mu$ in the bin $z_i$, it is not enough to measure clustering at $z_i$ and weak lensing at $(z_i,z_j)$. We need instead measurements of these quantities in a variety of redshifts larger than $z_i$. Two methods have been used to account for this fact. The simplest way is to assume a given evolution for $\mu$ with $a$, for example, see~\cite{DES:2018ufa,eBOSS:2020yzd}
\begin{align}
\mu(a)=1+\mu_0\frac{\Omega_\Lambda(a)}{\Omega_{\Lambda,0}}\, ,   
\end{align}
where only $\mu_0$ is a free parameter, that is constrained from the data. However, if $\mu$ does not evolve in this way, the constraints on $\mu_0$ are not valid. Another possibility is to parametrize $\mu$ and $\Sigma$ in terms of their values in a number of redshift nodes, and then interpolate between the nodes to obtain continuous functions, that can be used to solve Eq.~\eqref{eq:deltaevol}, see~\cite{Raveri:2021dbu}. This method does not assume any time evolution, but it depends on the chosen interpolation method, which introduces arbitrary correlations between the nodes. For example, using a cubic spline tends to suppress sharp changes in these functions~\cite{Raveri:2021dbu}. In contrast, in our approach, since no evolution equation needs to be solved, no interpolation is needed: $\hat{J}$ can be measured directly in each redshift bin. A theory prior on the evolution of $\hat{J}$ with redshift can be introduced if we want, for example, to reduce the number of free parameters, but this is not required. 

Finally, our approach has the advantage to be fully model-independent: we directly measure $\hat{J}$, i.e.\ the evolution of $\Phi+\Psi$, without any assumption on the theory of gravity or on the behavior of dark matter. This is not the case for the standard $\mu-\Sigma$ approach, which relies on the validity of Eq.~\eqref{eq:deltaevol} to find $\delta$ for a given $\mu$. As shown in~\cite{SvevaNastassiaCamille}, if dark matter couples differently than baryonic matter to gravity, or if dark matter is affected by a fifth (nongravitational) force~\cite{levon}, this equation is modified and the constraints on $\mu$ and $\Sigma$ are not valid.

Measuring $\hat{J}$ is therefore much more direct and robust than measuring $\Sigma$ and $\mu$. It is actually completely equivalent to measuring $\hat{f}=f\sigma_8$ from RSD. These measurements can indeed also be done in a completely model-independent way, in each of the redshift bins of the survey. In contrast, going from $f\sigma_8$ to $\mu$ requires to solve the evolution equation~\eqref{eq:deltaevol} for $\delta$. 
In the next section, we will present an example where measuring $\hat{J}$ is very useful and allows us to measure $\eta$ redshift bin by reshift bin, without having to assume anything on the evolution of $\mu$, and, as stated previously, without having to assume a specific behavior for dark matter. \itt{Note that from Table~\ref{tab:comp}, we expect the relative constraints on $\hat{J}$ and $\hat{f}$ to be of similar amplitude as the relative constraints on $\Sigma$ and $\mu$.}

To finish this section of comparison with other parametrizations, it is worth considering the recent analysis presented in~\cite{2022arXiv220603499C}. The author considers a template-fitting approach to constrain the growth of matter perturbations with cosmic shear analyses in a model-independent way, which is similar to our goal. There is however an important difference with respect to our approach, which is that in~\cite{2022arXiv220603499C}, GR is assumed, and therefore Eq.~\eqref{eq1} is used to relate the evolution of $\Phi+\Psi$ to the evolution of the density. The main goal of this method is therefore to measure the growth of density, without being affected by the bias. Hence, the free function that is fitted from the data is $\Omega_{\rm m}(z)\sigma_8(z)$. In our case, on the other hand, we use lensing to measure directly the evolution of $\Phi+\Psi$, i.e.\ $\hJ(z)$, without assuming GR. Note that because of that, we choose our reference power spectrum at $z=z_*$ (where we assume GR to be recovered) instead of $z=0$, as is done in~\cite{2022arXiv220603499C}. Furthermore, we assume the shape of the template power spectrum to be given by \itt{CMB} measurements, instead of assuming a fiducial cosmology and accounting for it with an Alcock-Paczy\'nski parameter. We also differ in the fact that we use mildly nonlinear scales with a GR nonlinear boost factor. Finally, another relevant difference is that in~\cite{2022arXiv220603499C} the author considers the cosmic shear observable and, because of this, he includes the BNT nulling technique proposed in~\cite{BNT}. Such technique allows the author to obtain localized cosmic shear kernels and therefore constrain the growth at different redshift bins. In our analysis, we consider the galaxy-galaxy lensing observable. Therefore, the agnostic $\hat{J}$ function only appears at the level of the lenses, which are already localized. 

Other recent model-independent analyses to test modified gravity that are worth mentioning are the Dark Energy Survey Year 1 analysis splitting between growth and geometry~\cite{DESY1split} and the Dark Energy Survey Year 3 analysis binning $\sigma_8$ as a function of redshift~\cite{DES:2022ygi}. The former considered a split of a subset of cosmological parameters, such that one parameter was sensitive to the growth of perturbations and the other one was sensitive to the geometry of the Universe. Within $\Lambda$CDM, the two parameters should agree and provide the same value. In the Year 3 analysis, a successor of this method was considered by introducing a set of amplitudes (one per redshift bin) that scale the linear power spectrum. These translate into the value of $\sigma_8$ at redshift 0 based on the amplitude of structure in a given redshift bin. These methods can be seen as consistency tests of $\Lambda$CDM, where any deviation would consist in an indication for beyond-$\Lambda$CDM physics. Instead, our method focuses on directly measuring the evolution of the gravitational potentials and the anisotropic stress as a function of redshift in a model-independent way.

\subsection{Gravitational redshift: \texorpdfstring{$\Psi$}{Lg}}

The method to measure directly the time distortion $\Psi$ has been presented in detail in~\cite{Sobral-Blanco:2022oel}. Here we simply summarize the main points. Following the same steps as in Sec.~\ref{sec:sum}, we introduce a new function $I(k,z)$ to encode the evolution of the gravitational potential $\Psi$ such that
\begin{align}
\label{eq:defI}
T_\Psi(k,z)=\frac{\HH^2(z)I(k,z)}{\HH^2(z_*)D_1(z_*)}\itt{\sqrt{B(k,z)}}T_\Psi(k,z_*)\, ,
\end{align}
\itt{where $B$ is the boost defined in Eq.~\eqref{eq:boost}. In our forecasts, we restrict the correlation function to separations larger than $d_{\rm min}$, chosen such that nonlinearities are negligible, i.e.,\ that the boost plays no role, see Sec.~\ref{sec:GRSKA}.}

As shown in~\cite{Sobral-Blanco:2022oel}, the evolution of $\Psi$ can be measured by cross-correlating the over-density, $\Delta$, of two populations of galaxies, e.g.,\ a bright and a faint population. $\Psi$ contributes to the galaxy over-density through the effect of gravitational redshift, which shifts to the red the spectrum of galaxies situated in a gravitational potential well. This effect adds to the density and RSD in Eq.~\eqref{eq:Delta} and generates a contribution of the form $\partial_r\Psi/\HH$ in $\Delta$~\cite{Bonvin:2014owa}. This term has the particularity to produce asymmetries in the distribution of galaxies~\cite{Bonvin:2013ogt}. Hence it was proposed to measure it by fitting for a dipole in the cross-correlation of bright and faint galaxies~\cite{McDonald:2009ud,Croft:2013taa,Bonvin:2013ogt}. However, since Doppler effects also contribute to such a dipole, one needs a method to disentangle the two types of contributions. More precisely, the dipole is sensitive to the following contributions in $\Delta$ (that are often called relativistic effects)~\footnote{Note that $\Delta^{\rm rel}$ contains other relativistic contributions, like Shapiro time-delay, integrated Sachs-Wolfe, and gravitational lensing~\cite{Bonvin2011,Yoo:2009au,Challinor:2011bk}, but these contributions are strongly subdominant and negligible at the scales and redshifts used in our analysis~\cite{Jelic-Cizmek:2020pkh,Euclid:2021rez}.} 
\begin{align}
\Delta^{\rm rel}(z,\bn)=&\frac{1}{\mathcal H}\partial_r\Psi+\frac{1}{\mathcal H}{\dot \bV}\cdot \mathbf n \label{eq:Delta_rel}\\
&+\left(1-5s+\frac{5s-2}{\mathcal H r}-\frac{{\dot{\HH}}}{\mathcal H^2}+f^{\rm evol}\right)\mathbf V\cdot \mathbf n\, ,\nonumber
\end{align}
where a dot denotes derivative with respect to conformal time, $s$ is the magnification bias and $f^{\rm evol}$ is the evolution bias.

In~\cite{Sobral-Blanco:2022oel}, it was shown that by combining measurements of the monopole, quadrupole, and hexadecapole of the correlation function, with a measurement of the dipole, one can measure separately the evolution of $\Psi$ and the evolution of the velocity, in a model-independent way. More precisely it was shown that the quantity $\hat{I}$ given by
\begin{align}
\hat{I}(z)\equiv\frac{I(z)\sigma_8(z)}{D_1(z)}=\frac{I(z)\sigma_8(z_*)}{D_1(z_*)} \label{eq:hatI}
\end{align}
can be measured directly, redshift bin by redshift bin. Note that as before, we neglect here the $k$-dependence of $\hat{I}$, since it is a good approximation in the quasi-static limit.

The parameter $\hat{I}$ is a new parameter, that adds to the ones defined in Table~\ref{tab:comp}. If galaxies obey Euler's equation, which is the case if the weak equivalence principle for dark matter and baryonic matter is valid, then $\hat{I}$ can be related to $\hat{f}$ through 
\begin{align}
 \hat{I}=\frac{2\hat{f}}{3}\left[\frac{\HH'}{\HH}+\frac{\hat{f}'}{\hat{f}}+1 \right]\,.
 \label{eq:Ifromf}
 \end{align}

It is therefore possible to reconstruct the evolution of $\Psi$ from RSD, as has been shown in~\cite{Motta:2013cwa}. On the other hand, if Euler's equation is not valid, for example if dark matter is sensitive to a fifth force, or if baryons and dark matter are not coupled in the same way to gravity, then Eq.~\eqref{eq:Ifromf} is not valid~\cite{levon}. In this case, $\hat{I}$ has to be considered as an independent function, that cannot be inferred from RSD measurements. The dipole is therefore an important new observable, since it will allow us to measure $\hat{I}$ directly, without having to assume anything on the behavior of dark matter. \itt{With current surveys, the dipole is unfortunately not detectable~\cite{Gaztanaga:2015jrs}. Forecasts show however that the coming generation of surveys, like DESI and SKA2, will be able to measure it robustly~\cite{Bonvin:2018ckp,Beutler:2020evf,Saga:2021jrh}.}

\subsection{Anisotropic stress}

Now that we have derived the parametrization of the gravitational lensing and gravitational redshifit observables, we can combine them to measure $\eta=\Phi/\Psi$. We have
\begin{equation}\label{eq13}
    \frac{\hat{I}(z)}{\hJ(z)}=\frac{T_{\Phi+\Psi}(k,z_*)}{T_{\Psi}(k,z_*)}\frac{T_{\Psi}(k,z)}{T_{\Phi+\Psi}(k,z)}=\frac{2\Psi}{\Phi+\Psi}=\frac{2}{1+\eta}\,,
\end{equation}
\itt{where in the first equality we use Eqs.~\eqref{eq:evolJ} and~\eqref{eq:defI}. The boost, $B(k,z)$, cancels in the ratio since it affects in the same way the evolution of $\Psi$ and of $\Phi+\Psi$ (it encodes indeed the nonlinear evolution of matter density -- see Eq.\,\eqref{eq:boost}). Moreover, in the second equality we use} that $(\Phi+\Psi)(z_*)=2\Psi(z_*)$, since GR is recovered at $z_*$.

We note that we consider the ratio of $\hat{I}$ over $\hJ$, or equivalently the ratio of $\Psi$ over the Weyl potential, instead of the inverse, because our constraints on $\Psi$ are weaker, as we will see in the following sections. The fact of having weaker constraints allows $\Psi$ to become compatible with a null value, which would introduce numerical instabilities in the inverse of Eq.\,(\ref{eq13}).

Equation~\eqref{eq13} is a novel estimator of the anisotropic stress, $\eta$, which is fully model-independent. It is directly built from measurements of the functions $\hat{J}$ and $\hat{I}$ in the bins of the surveys. If the ratio $\hat{I}/\hat{J}$ differs from 1, we can then unambiguously conclude that gravity is modified. In contrast, the standard $\mu-\Sigma$ approach does not allow us to measure $\eta$ in a model-independent way. In particular, with the standard approach, we could detect an apparent deviation from GR: $\eta\neq 1$, even if gravity is not modified and $\Psi=\Phi$. As has been discussed in~\cite{levon}, this is due to the fact that if dark matter is affected by a fifth force, RSD do not provide a measurement of the true $\mu$. As a consequence, the observed $\eta$, which is inferred from a measurement of $\Sigma$ and $\mu$, will not be the true $\eta$. Our method is therefore crucial if we want to regard $\eta\neq 1$ as a smoking gun for modified gravity.

\section{Methodology} 
\label{sec:methodology}

In this section we present the methodology used to forecast the constraints on the observables presented in Sec.~\ref{sec:modeling}. We first describe the galaxy surveys considered and their settings, and then show the Fisher matrix forecast used in this work.

\begin{figure}
    \centering
    \includegraphics[width=\linewidth]{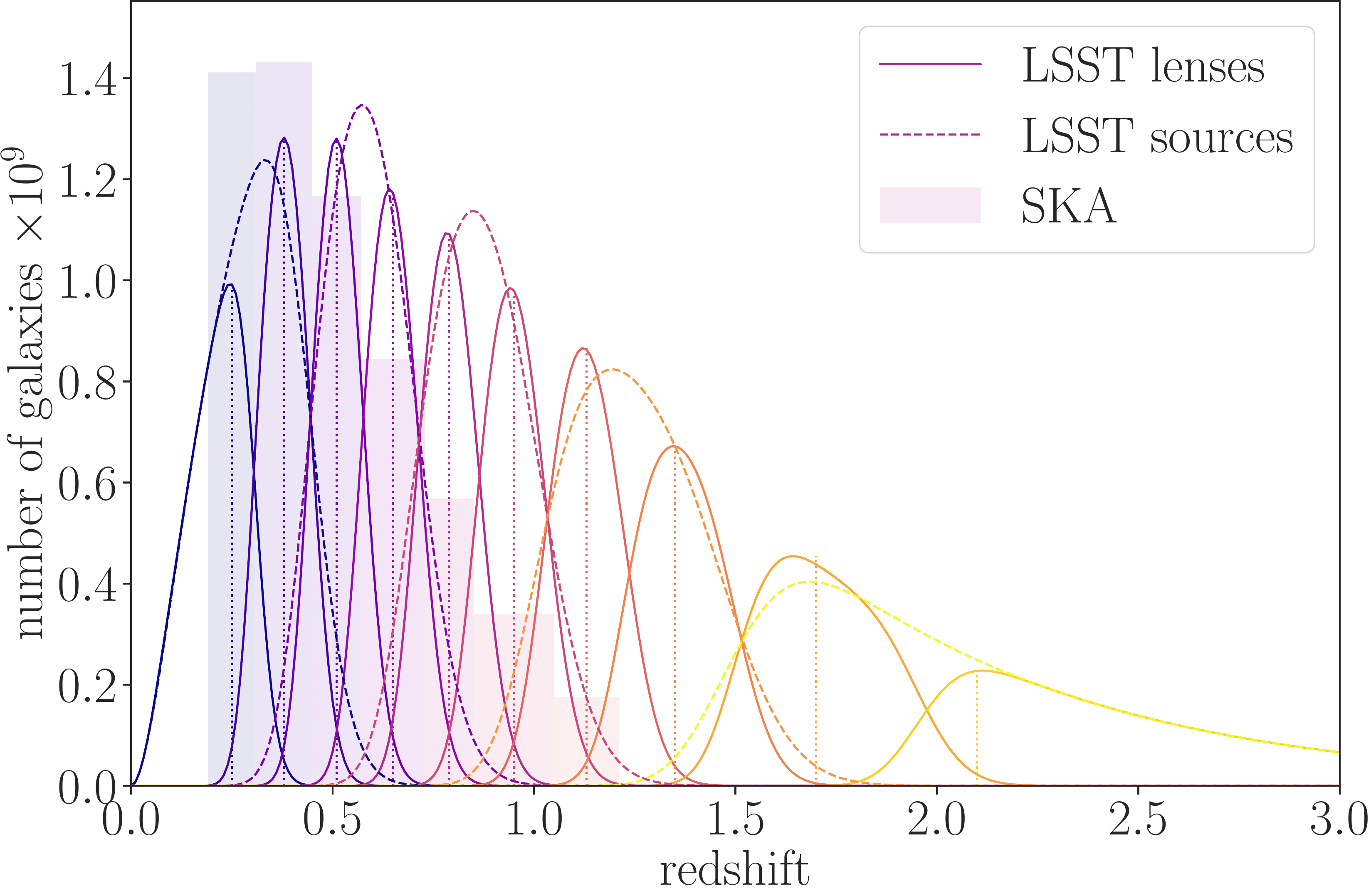}
    \caption{Number of galaxies as a function of redshift for the different samples considered. The solid lines stand for the LSST lenses, the dashed lines represent the LSST sources, and the bars represent the SKA spectroscopic top-hat bins, that have been adapted to the LSST lenses. The effective redshift for each tomographic bin is represented with a vertical dotted line.}
    \label{fig1}
\end{figure}

\subsection{Gravitational lensing with LSST}
In the following years we will have access to a huge amount of exquisite photometric data from the future Stage-IV galaxy surveys. Some examples are the Vera C. Rubin Observatory that will carry out the Legacy Survey of Space and Time (LSST\,\footnote{\url{https://www.lsst.org}},~\cite{2019ApJ...873..111I}), the Euclid satellite\,\footnote{\url{https://www.euclid-ec.org}}~\cite{2011arXiv1110.3193L}, or the Nancy G. Roman Space Telescope\,\footnote{\url{https://roman.gsfc.nasa.gov}}~\cite{2015arXiv150303757S}. In this work we focus on LSST, whose main science objective from a cosmological perspective is to probe dark energy and dark matter. In order to achieve this, the baseline survey will cover 18\,000 square degrees of the Southern sky during 10 years to obtain accurate photometry in multiple optical bands. Such a survey will allow us to obtain accurate photometric redshifts and weak lensing shear measurements for about 27 galaxies per arcmin$^2$. This will enable precise photometric galaxy clustering, cosmic shear, and galaxy-galaxy lensing cosmological analyses, also called 3x2pt analyses.

In more detail, we consider the survey settings provided with the public version of \texttt{CosmoSIS}\,\cite{2015A&C....12...45Z}. These consist in five equipopulated tomographic bins for the sources and ten equipopulated tomographic bins for the lenses with a total of 27 galaxies per arcmin$^2$, for both sources and lenses. We represent the galaxy distributions in Fig.\,\ref{fig1}. Additionally, we consider a linear galaxy bias model with a constant fiducial set to $b=2$, but treated as a nuisance parameter in each tomographic bin and marginalized over\,\footnote{We note that a realistic galaxy bias might have some redshift dependence. However, since we marginalize over its value at each redshift bin, we have verified that a different fiducial value for the galaxy bias does not change the size of the final constraints. We therefore keep $b=2$ as fiducial, for simplicity.}. We also consider the nonlinear alignment model for intrinsic alignments\,\cite{Bridle_2007,10.1111/j.1365-2966.2012.21099.x} with a fiducial amplitude set to $\itt{A_{\rm IA}=1}$ but allowed to vary. We note that we limit our analysis to multipoles between $\ell=20$ and $\ell=2627$ in the optimistic scenario and between $\ell=20$ and $\ell=750$ in the pessimistic scenario. In both cases we do not go too much into the nonlinear regime, and therefore the linear modeling for galaxy bias and intrinsic alignment is still a good approximation for our forecasts. Finally, we consider an ellipticity total dispersion of $\sigma_{\epsilon}=0.3$.

It is important to mention that \itt{we assume GR is valid at small scales where intrinsic alignments are important, that is, we include the intrinsic alignment contribution using a standard GR modeling, even when constraining $\hat{J}$. The main reason for this choice is the lack of intrinsic alignment models for modified gravity theories. However, we marginalize over the amplitude of this effect to account for its impact. We} neglect some observational systematic uncertainties, like a shear calibration bias, or biases in the mean of our galaxy distributions. Moreover, we neglect some contributions to the signal like the impact of magnification or RSD for photometric galaxy clustering and galaxy-galaxy lensing observables. This is also the approach followed in~\cite{ISTFpaper}, for example. Note that we do not expect magnification to alter the measurement of $\hat{b}$ and $\hat{J}$ at small redshift, where its contribution to the signal is strongly subdominant. Since, as we will see, $\eta$ is only well measured at small redshift with our estimator, our analysis should be independent of this contamination. Also RSD are partially washed out in thick tomographic bins and therefore a subdominant contribution to the total signal. In any case, this analysis focuses on determining whether a model-independent measurement of the anisotropic stress can be performed with future observations. Therefore, our first aim is to check its feasibility in an optimistic scenario where the main systematic uncertainties are under control. Obviously this analysis will need to account for observational uncertainties once real observations are available and to obtain more precise predictions.

\subsection{Gravitational redshift with SKA}
\label{sec:GRSKA}

Future Stage-IV galaxy surveys will also provide us with precise spectroscopic data probing the galaxy clustering in our Universe. An example of such a survey is the phase 2 of the future Square Kilometer Array (SKA)\,\footnote{\url{https://www.skatelescope.org}}, with which we will be able to observe close to a billion of galaxies between redshifts $z=0.1$ and $z=2.0$. It is expected that this new data will drastically improve our measurements of the growth history, achieving subpercent measurements of $f(z)\sigma_8(z)$ \cite{2016ApJ...817...26B} using the even multipoles of the galaxy clustering correlation function. Furthermore, by combining the measurement of the even multipoles with the dipole we will be able to measure the gravitational redshift with a precision of $10-30\%$ at late times (see \cite{Sobral-Blanco:2022oel}). In this work, we use this measurement to build an estimator for the anisotropic stress by comparing the functions $\hat{I}(z)$ and $\hat{J}(z)$ of Eq.~\eqref{eq:hatI} and Eq.\,(\ref{eq11}), respectively. Importantly, we adapt the redshift bins considered in~\cite{Sobral-Blanco:2022oel} for $\hat{I}(z)$, in order to obtain constraints at the same effective redshifts than those for $\hat{J}(z)$. This can easily be done for SKA2, thanks to the very precise determination of redshift. The new top-hat galaxy distributions and the effective redshifts are shown in Fig.\,\ref{fig1}. 

We consider the number density and volume specifications presented in~\cite{2016ApJ...817...26B} and follow the approach used in~\cite{Sobral-Blanco:2022oel} to split the populations of galaxies between faint and bright such that we have the same number of each luminosity type per redshift bin. For the fiducial of the galaxy biases we use the exponential fitting functions
\begin{align}
    b_{\rm B} &= ce^{dz}+\frac{\Delta b}{2}\,,\\
    b_{\rm F} &= ce^{dz}-\frac{\Delta b}{2}\,,
\end{align}
for the bright and faint populations, respectively, where $c=0.554$ and $d=0.783$, following~\cite{2016ApJ...817...26B}. As in~\cite{Sobral-Blanco:2022oel}, we assume a difference between the two galaxy biases of $\Delta b=1$. This is consistent with the $\mathcal{O}(1)$ difference that has been measured for BOSS in~\cite{Gaztanaga:2015jrs}. In our forecasts, we then treat $b_{\rm B}$ and $b_{\rm F}$ as free parameters in each redshift bin and marginalize over them. For the magnification bias, we use the model developed in ~\cite{SvevaNastassiaCamille}, and we neglect the evolution bias for the two populations. Once data will be available, both the evolution bias and the magnification bias of the two populations will be directly measurable from the average galaxy distribution. Therefore we keep these values fixed in our forecasts. 

Finally, we shall perform two forecasts by choosing different values for the minimum separations between galaxies. This $d_{\rm min}$ must be set at the scale on which the linear regime is a good approximation. Following~\cite{2020JCAP...08..004B}, for an optimistic scenario, we choose $d_{\rm min}=20\,{\rm Mpc}/h$, while for the pessimistic case we set $d_{\rm min}=32\,{\rm Mpc}/h$. \itt{In this regime, the boost in Eq.~\eqref{eq:defI} plays a negligible role and can therefore be set to 1 in our forecasts.} In addition, we set the maximum separation such that it is consistent with the size of our smallest redshift bin, $d_{\rm max}=120\,{\rm Mpc}/h$, in all of our bins. 

\subsection{Fisher matrix forecasts}

Once we have our theoretical predictions for the observables derived in Sec.\,\ref{sec:modeling} with the specifications provided above, we can forecast the uncertainties on the different parameters using a Fisher matrix formalism. \itt{We summarize the parameters considered in our analyis, together with their fiducial values, in Tables\,\ref{tab:fids_fixed} and \ref{tab:fids} in the Appendix, but for simplicity our vector of free parameters is given by} 
\itt{\begin{equation}
    \theta=\{
    \hat{J}_i,\hat{b}_i,A_{\rm IA},\hat{I}_j,\hat{f}_j,\hat{b}_{\text{B},j},\hat{b}_{\text{F},j}\}\,,
\end{equation}}
\itt{where $i$ runs over the ten tomographic bins for LSST and $j$ runs over the seven bins for SKA2.}

We recall that the Fisher matrix is defined as the expectation value of the second derivative of the logarithm of the likelihood with respect to the parameters of the model:
\begin{equation}
    F_{\alpha\beta}=\Braket{-\frac{\partial^2\,\text{ln}(\mathcal{L})}{\partial \theta_{\alpha}\partial \theta_{\beta}}}\,.
\end{equation}

For a Gaussian likelihood, and neglecting any dependence of the covariance of the observables on the model parameters, the Fisher matrix can be expressed as
\begin{equation}
    F_{\alpha\beta}=\sum_{pq}\frac{\partial \mu_p}{\partial \theta_{\alpha}}\left(\text{C}^{-1}\right)_{pq}\frac{\partial \mu_q}{\partial \theta_{\beta}}\,,
\end{equation}
where $\mu$ is the mean of the data vector and $\text{C}$ is the covariance matrix of the data. These will correspond to the galaxy-galaxy lensing and galaxy clustering harmonic spectra for LSST, and to the correlation function multipoles for SKA2. We consider a Gaussian covariance for the former, meaning that we account for the cosmic variance and shape/shot noise, but neglect non-Gaussian terms like the supersample covariance, as was done in~\cite{ISTFpaper}. For SKA2, we include shot noise and cosmic variance in the variance of the multipoles (see Appendix~C of~\cite{Bonvin:2018ckp}) and account for cross-correlations between the different combinations of luminosity pairs. \itt{We note that, because of our $\Lambda$CDM fiducial, we use GR to compute these covariances.}

Once the Fisher matrix is computed, we estimate the covariance matrix of the model parameters as the inverse of the Fisher matrix:
\begin{equation}
C_{\alpha\beta}=\left(\text{F}^{-1}\right)_{\alpha\beta}\,.
\end{equation}

In practice, we use \texttt{CosmoSIS} to call the \texttt{CAMB} Boltzmann solver\,\cite{2000ApJ...538..473L,2012JCAP...04..027H} and compute the Fisher matrix for the $\hJ$ parameters. We then use the same Boltzmann solver to build the Fisher matrix for the $\hat{I}$ parameters from the correlation function multipoles. We note that our gravitational lensing observable is also sensitive to $\hb$ and the amplitude of intrinsic alignments, while our gravitational redshift observable depends on the generic growth function $\hat{f}=f\sigma_8$~\footnote{Note that we treat $\hat{f}$ as a free function, with unknown time evolution. In particular, we do not assume that $f=\frac{d\ln D_1}{d\ln a}$, since this is only true if the continuity equation for dark matter is valid. Since we want to remain agnostic about the behavior of dark matter, we do not assume that this equation is valid.} and galaxy biases $\hat{b}_{\rm B}=b_{\rm B}\sigma_8$ and $\hat{b}_{\rm F}=b_{\rm F}\sigma_8$. These additional parameters are considered nuisance parameters and we marginalize over them when providing constraints on $\hJ$ or $\hat{I}$. We also remind the reader that these observables depend on the cosmological parameters providing the spectrum of matter perturbations at $z=z_*$ but we consider the cosmology fixed.

In addition to the forecast constraints on $\hJ$ and $\hat{I}$, sensitive to $\Phi+\Psi$ and $\Psi$, respectively, we want to combine them to constrain the anisotropic stress. Instead of building a Jacobian transformation to move from one set of parameters to another one, and to keep the nonlinearities that may arise in the transformation, we generate synthetic chains from the individual Fisher matrices. In more detail, given a Fisher matrix for the $\hJ$ parameters and a Fisher matrix for the $\hat{I}$ parameters, we invert them to obtain the covariances of the parameters. Then, for each one of them, we generate a mock chain centered at our fiducial and with random points drawn from a multidimensional Gaussian distribution with the corresponding covariance. We finally generate a third chain from the ratio of the other two at each point, which provides the posterior on $\hat{I}/\hJ=2/(1+\eta)$. It is important to mention that we only consider seven redshift bins for $\hat{I}$, while we use ten redshift bins for $\hJ$, as can be seen in Fig.\,\ref{fig1}. The main reason for this choice is the lack of constraining power on $\hat{I}$ at higher redshifts. Therefore, we further marginalize over the last three tomographic bins for the gravitational lensing observable when it is combined with the gravitational redshift.

\section{Results} 
\label{sec:results}

In this section we present the main results of the analysis. We first focus on the gravitational lensing observable and the constraints on $\Phi+\Psi$. We then present the results on $\Psi$ coming from the dipole of the correlation function. Finally, we derive constraints on the anisotropic stress from the combination of the two observables.

\subsection{Gravitational lensing}

\begin{figure}
    \centering
    \includegraphics[width=\linewidth]{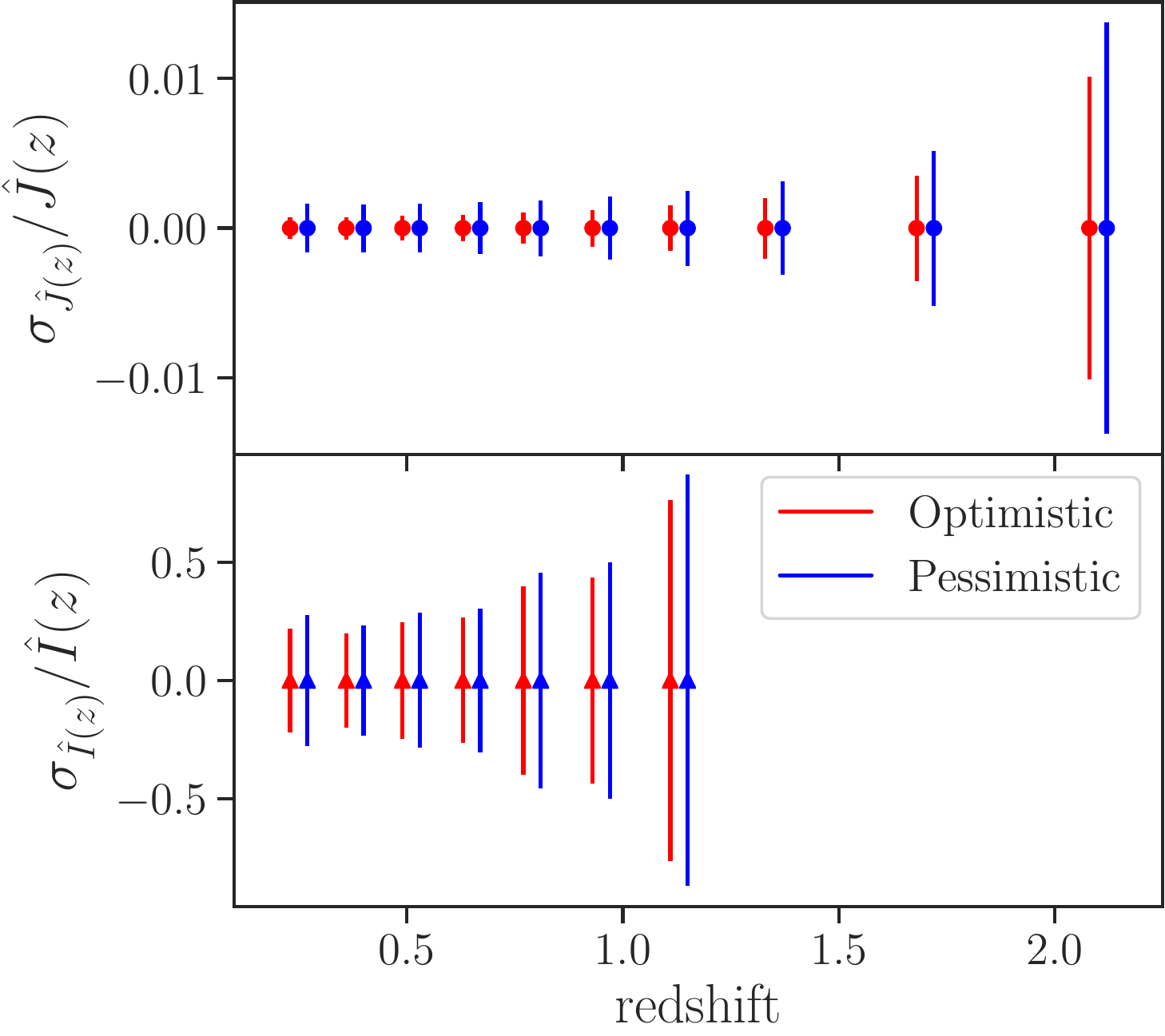}
    \caption{Forecast $1\,\sigma$ uncertainties for $\hJ$ (top panel) and $\hat{I}$ (bottom panel) with respect to their fiducial value for each tomographic bin. The red \itt{(light gray)} error bars correspond to the optimistic scenario, while the blue \itt{(dark gray)} error bars represent the pessimistic settings described in the text.}
    \label{fig1p5}
\end{figure}
In the top panel of Fig.\,\ref{fig1p5} we present the $1\,\sigma$ forecast uncertainties for $\hJ$ with respect to their fiducial value as a function of redshift. We show the optimistic scenario in red, corresponding to scale cuts $\ell_{\rm max}=2627$, and the pessimistic scenario in blue, corresponding to $\ell_{\rm max}=750$. A given offset has been added in the x axis for illustrative purposes.

As can be seen in Fig.\,\ref{fig1p5}, there is a degradation of the constraints on $\hJ$ as a function of redshift. This behavior is essentially due to two effects. First, at high redshift, the tomographic bins are wider, which implies that there is a more significant smoothing of the galaxy clustering distribution along the line-of-sight. Because of this, the amplitude of the galaxy-galaxy lensing spectra decreases, leading to worse constraints on $\hJ$. Second, at high redshift, the lenses are necessarily closer to the sources, which decreases the lensing kernel, and in addition, the lenses are correlated with a smaller number of bins, which decreases the number of independent measurements of $\hat{J}$ at that redshift. Let us for example consider the tomographic bin number two for the lenses in Fig.\,\ref{fig1}, which is centered at redshift $z\sim 0.4$. 
The lensing efficiency will peak at roughly the double of this redshift, implying that the sources in the third tomographic bin, which is centered at redshift $z\sim 0.9$, will provide a high signal-to-noise measurement. In addition, this second bin is also correlated (albeit less strongly) with the bins number two, four, and five of the sources, providing four independent measurements of $\hat{J}$ at $z\sim 0.4$. On the other hand, if we consider the high-redshift lenses, like tomographic bins number seven or higher (effective redshifts higher than $z\sim 1.2$),  
there are no sources at the double of these effective redshifts, where the signal would be the strongest. This means that only the very low-redshift tail of the lenses distributions will be close to half the effective redshift of the sources. Therefore, this will lead to a decrease of the lensing kernel and thus a decrease of the galaxy-galaxy lensing signal. Furthermore, since the lenses are not correlated with the sources at lower redshift, the lenses bin number seven  
is only correlated with the sources bin number four and five, which provides only two independent measurements of $\hat{J}$. All together, these effects lead to worse constraints on $\hJ$ at high redshift. We note that all tomographic bins have the same number of galaxies with the same ellipticity dispersion. Therefore, the shape and shot noise are the same for all redshifts. Cosmic variance is instead a bit smaller at high redshift, given the larger volume, but not enough to compensate the effects mentioned above. However, even accounting for the degradation as a function of redshift, LSST will be able to constrain $\hJ$ as a function of redshift at \itt{less than} percent level, and therefore probe the Weyl potential with very high precision.

Let us mention that we have validated our method to measure $\hat{J}$, by comparing it with the standard method, assuming $\Lambda$CDM. For this we have proceeded in the following way. First, we have verified that replacing the standard $\Lambda$CDM harmonic power spectra by the ones provided in \itt{Eqs.\,(\ref{eq:DeltakappaJ},\ref{eq:DeltaDeltaJ})} with the $\hat{J}$ and $\hat{b}$ values in Eqs.\,(\ref{eq11}-\ref{eq14}), we recover the same constraints on the cosmological parameters at the level of $0.1\,\%$. We note that in this test we have fixed the values of $\hat{J}$ and $\hat{b}$ and constrained the cosmological parameters, with the goal of testing the implementation of the new harmonic spectra against the standard method. In a second step, we have assumed $\hat{J}$ and $\hat{b}$ to be constant within each redshift bin and moved them out of the integral. Their values have then been fixed according to Eqs.\,(\ref{eq11}-\ref{eq14}) at the effective redshift of the corresponding bin. Under this assumption, the recovered constraints on the cosmological parameters \itt{degrade by a factor between 1 and 3.5 compared to the constraints obtained} with the standard approach. 
Such a discrepancy is expected, since by fixing the value of $\hat{J}$ and $\hat{b}$ in each redshift bin, we remove the information coming from the evolution of these quantities inside the bins. Performing the same test with twice the number of redshift bins, we have found that the \itt{degradation} 
reduces to \itt{a factor between 1 and 1.8}, 
showing that as we increase the number of bins and the approximation of constant $\hat{J}$ and $\hat{b}$ is more valid, we recover the standard constraints with our new implementation. 

Besides validating our methodology, this test also shows the level of degradation on the cosmological constraints due to our requirement of model-independence. Since we want to measure the evolution of the Weyl potential without assuming a specific theory of gravity, we are limited by the size of the tomographic bins. \itt{Contrary to standard methods, where the evolution within a redshift bin is given by the model, here we can only measure the value of the Weyl potential at the effective redshift of the bins, therefore loosing part of the information.} 

\subsection{Gravitational redshift}
Let us now focus on the spectroscopic side of the analysis with the dipole measurements. We present the main results in the bottom panel of Fig.\,\ref{fig1p5}. As for $\hJ$ in the top panel, we present the $1\,\sigma$ forecast uncertainty on $\hat{I}$ with respect to their fiducial value as a function of redshift. We present both the optimistic (red) and pessimistic (blue) scenarios, which correspond to $d_{\rm min}=20\,\text{Mpc}/h$ and $d_{\rm min}=32\,\text{Mpc}/h$, respectively. As for the gravitational lensing observable, the uncertainties on $\hat{I}$ increase as a function of redshift. This is due to the fact that the signal decreases with increasing redshift, given the decreasing growth at higher redshifts.  
Moreover, shot-noise increases quickly as a function of redshift, as can be seen in the drop of the number of galaxies in Fig.\,\ref{fig1}.  

Overall, SKA2 will be able to constrain $\hat{I}$ as a function of redshift at the level of $\sim 20\%$. These results are consistent with the forecasts presented in~\cite{Sobral-Blanco:2022oel}, although different redshift bins have been considered in this work. It is important to note that the constraining power on $\hJ$ is much stronger than on $\hat{I}$, but this is an expected result. The signal-to-noise ratio (SNR) of the galaxy-galaxy lensing observable is already of 148 with current observations~\cite{2022PhRvD.105h3528P}, while the expected SNR for gravitational redshift is much lower. Using the bins defined in this analysis, the SNR can be computed as
\begin{align}
 {\rm SNR}^{\Psi}=\sum_z\sum_{ij}\xi^{\Psi}_1(d_i,z){\rm cov}^{-1}(d_i,d_j,z)\xi^{\Psi}_1(d_j,z)\,  ,
\end{align}
where $\xi^{\Psi}_1(d_i,z)$ is the gravitational redshift contribution to the dipole, i.e.\ the contribution from the first term in Eq.~\eqref{eq:Delta_rel}, evaluated at separation $d_i$ and in the redshift bin $z$, and cov$(d_i,d_j,z)$ stands for its covariance at separations $d_i$ and $d_j$ and redshift bin $z$. Summing from 20 to 120\,Mpc$/h$, we find a total SNR of 8. Note that the SNR of the dipole is significantly larger, due to the Doppler effects, but these terms do not contribute to the constraints on $\hat{I}$. Therefore, it is not surprising that the constraints obtained with gravitational lensing will be much more stringent. However, the particularity of the dipole is that it allows us to directly constrain $\hat{I}$ and therefore directly probe $\Psi$ in a model-independent way, something that cannot be done with any other observable at cosmological scales.

\itt{Finally, let us stress that the constraining power on $\hat{I}$ \itt{directly} depends on the magnitude of the bias difference \itt{between the bright and the faint population. In} this work, \itt{we assumed a bias difference of 1, consistent with what} has been measured for BOSS in \cite{Gaztanaga:2015jrs}. \itt{If the bias difference turns out to be smaller (see, e.g.,} the analysis of \cite{alam}, \itt{which finds} a bias difference of order $\mathcal{O}(0.5)$), the constraints presented in this work worsen by a factor of $\sim 2$. \itt{However, the final results still provide stringent constraints.}}

\subsection{Anisotropic stress with the combination of probes}

\begin{figure}
    \centering
    \includegraphics[width=\linewidth]{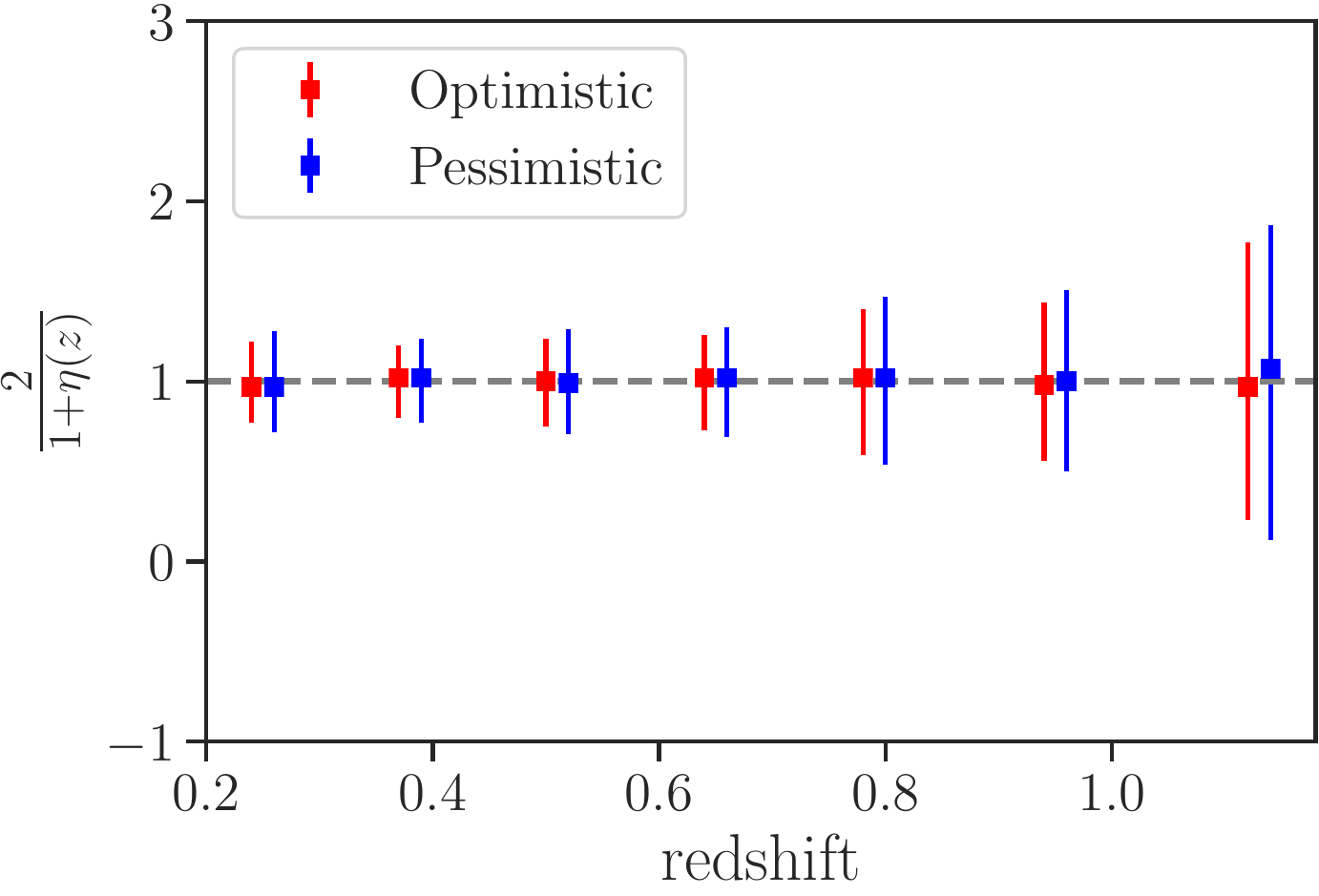}
    \caption{Reconstruction of $2/(1+\eta)$ as a function of redshift. The dashed horizontal line represents the fiducial, while the vertical error bars show the optimistic (red\itt{, light gray}) and pessimistic (blue\itt{, dark gray}) forecast uncertainties.} 
    \label{fig3}
\end{figure}

After computing the forecast constraints for both $\hJ$ and $\hat{I}$, we can combine them and place constraints on the anisotropic stress $\eta$. Following the methodology described in Sec.~\ref{sec:methodology} and after marginalizing over the last three tomographic bins for $\hJ$, we generate a mock chain of 50\,000 points for $\hJ$ and another one for $\hat{I}$. In each point we compute the ratio presented in Eq.\,(\ref{eq13}) and compute the posterior at each redshift bin. We present in Fig.\,\ref{fig3} the $1\,\sigma$ uncertainties on $2/(1+\eta)$ as a function of redshift for both the optimistic (red) and pessimistic (blue) scenarios. Again, an offset in the $x$ axis has been included for illustrative purposes. 

As it was the case for $\hJ$ and $\hat{I}$, there is a degradation of the constraining power as a function of redshift, which is given by the same physical effects since we are computing the ratio of the two quantities. We can appreciate a more significant degradation compared to the $\hJ$ case, but this is due to the fact that the uncertainties are driven by the uncertainties on $\hat{I}$. Overall, the combination of LSST and SKA2 will allow us to constrain the anisotropic stress, through the relation $2/(1+\eta)$, in a model-independent way as a function of redshift and at the level of $\sim20\,\%$ in the optimistic scenario ($\sim30\,\%$ in the pessimistic case). \itt{Note again that \itt{if 
the bias difference 
is of} order $\mathcal{O}(0.5)$, the final constraints on $\eta$ worsen by a factor of $\sim 2$.} Finally, let us mention that since the uncertainties are essentially dominated by the dipole, we do not expect a significant gain on the constraints if we add cosmic shear into the analysis. 

\section{Conclusions} 
\label{sec:conclusions}

In this work we have addressed the question of how well will future surveys constrain the laws of gravity. Given how open this question is, one way to answer it is by comparing the two gravitational potentials encoding the time distortion and the spatial distortion, that is $\Psi$ and $\Phi$, respectively.
This comparison is a key test of the validity of GR, since it compares directly the two independent degrees of freedom of the Universe's geometry. In contrast, the growth of structure is less direct: a deviation from the $\Lambda$CDM prediction can indeed be generated not only by modified gravity, but also by a fifth force acting on dark matter, or by a dark energy component that is clustering. Because of this, measuring $\eta\neq 1$ is often regarded as the smoking gun of modified gravity.

Many analyses exist in the literature constraining the anisotropic stress, $\eta$, with current observations~\cite{DES:2018ufa,eBOSS:2020yzd,DES:2022ygi,Noller:2020afd}, or forecasting the expected constraints with future surveys~\cite{Amendola:2012ky,Motta:2013cwa,Pinho:2018unz}. However, these analyses all rely on the validity of the weak equivalence principle. Hence, they are simply not valid if dark matter obeys a fifth force. In this work we have built a new estimator, which does not rely on the behavior (or even on the existence) of dark matter, and we have
forecast the expected constraints from future galaxy surveys.

Our estimator improves the way gravity is tested in two ways. First, it relies on a novel parametrization that we have developed for the galaxy-galaxy lensing observable, which allows us to directly measure the evolution of the Weyl potential, encoded in a free function $\hat{J}$, in each tomographic bin. We have shown that combining the galaxy-galaxy lensing with the galaxy clustering of the photometrically-detected galaxies allows us to break the degeneracy between this new function and the galaxy bias. We have then forecast the constraining capability of LSST with a Fisher matrix formalism and determined the uncertainties on $\hat{J}$ as a function of redshift (see Fig.\,\ref{fig1p5}). We have observed an expected degradation of the constraining power toward higher redshifts, but the overall constraints are at \itt{less than} percent level, showing that LSST will be able to provide very precise model-independent constraints of the late-time growth of  the Weyl potential. This novel parametrization has the strong advantage of being fully model-independent, contrary to the standard $\mu-\Sigma$ parametrization, which relies on the validity of the weak equivalence principle to measure $\mu$ and $\Sigma$ from weak lensing observables~\cite{SvevaNastassiaCamille, levon}.

The second improvement with respect to standard analyses is that our estimator is built from a direct measurement of the evolution of $\Psi$ (encoded in a new function $\hat{I}$) from the dipole of galaxy clustering. Using the Fisher matrix formalism, we have forecast the constraining power of a future survey like SKA2 on $\hat{I}$ as a function of redshift (see Fig.\,\ref{fig1p5}). We have observed an expected degradation as a function of redshift and, overall, significantly weaker constraints compared to $\hat{J}$. This is also expected given the difficulty to measure the relativistic dipole of the correlation function. Nevertheless, SKA2 will be able to put constraints of the order of $20\,\%$ 
on $\hat{I}$, therefore directly constraining the gravitational potential $\Psi$. These results are consistent with those presented in~\cite{Sobral-Blanco:2022oel}, although a different redshift binning has been considered in this work. This methodology has, again, the strong advantage of being model-independent. Current measurements of $\Psi$ are instead performed through RSD, assuming that the equivalence principle is valid. More precisely, $\Psi$ is inferred from the velocity, using Euler's equation.

Finally, we have combined the constraints on $\hat{J}$ from LSST and the constraints on $\hat{I}$ from SKA2 to constrain the anisotropic stress. More precisely, we have considered the combination $2/(1+\eta)$ and shown that these two future galaxy surveys will be able to constrain it at the $\sim 20\,\%$ level in a direct and model-independent way. This level of precision is similar to current constraints on $\eta$ obtained through the $\mu-\Sigma$ parametrization. The latest DES analysis does indeed constrain $\mu_0$ with a precision of 20\% and $\Sigma_0$ with a precision of 5\%~\cite{DES:2022ygi}, leading to a derived constraint on $\eta_0$ of the order of 25\%. However, as explained before, these constraints are not valid if dark matter obeys a fifth force. Moreover, they assume a given time evolution for $\mu$ and $\Sigma$, meaning that only $\eta_0$, i.e.\ the value of $\eta$ today can be constrained with this method. If this time evolution is not correct, then the constraints that are obtained are invalid. In contrast, in our case, $\eta$ is measured independently in each of the redshift bins of the surveys.

To conclude, let us mention that our analysis has considered a simple Fisher matrix formalism, accounting for some astrophysical systematic uncertainties, like galaxy bias or intrinsic alignments. However, an analysis using real observations should go beyond this proof-of-concept and account for observational systematic uncertainties, that we have assumed here to be under control.

\paragraph*{Acknowledgements.} This project has received funding from the Swiss National Science Foundation and from the European Research Council (ERC) under the European Union’s Horizon 2020 research and innovation program (Grant agreement No.~863929; project title ``Testing the law of gravity with novel large-scale structure observables").


\bibliographystyle{apsrev4-1}
\bibliography{main.bib}

\section*{Appendix A}
\itt{In this appendix we provide the fiducial values for all the parameters considered in our analysis, including those that have been fixed. The latter are presented in Table\,\ref{tab:fids_fixed}, while the former are shown in Table\,\ref{tab:fids}.}

\begin{table}[H]
\centering
\caption{\itt{List of fixed parameters considered in the Fisher analysis, together with their fiducial values in the $\Lambda$CDM model.}}
\label{tab:fids_fixed}
\begin{tabular}
{C{1.8cm}|C{6.5cm}}
\hline
Parameter  & Fiducial value \\ 
\hline
$\Omega_{\rm m,0}$ & $0.3111$\\ 
$\Omega_{\rm b,0}$ & $0.0490$\\ 
$h$ & $0.677$\\ 
$n_{\rm s}$ & $0.9665$\\ 
$A_{\rm s}$ & $2.105\times 10^{-9}$\\ 
$s_{\text{B}, 1}$ & $s_{\text{B}}(z=0.25) = 0.3706$ \\
$s_{\text{B}, 2}$ & $s_{\text{B}}(z=0.38) = 0.4665$ \\
$s_{\text{B}, 3}$ & $s_{\text{B}}(z=0.51) = 0.5757$ \\
$s_{\text{B}, 4}$ & $s_{\text{B}}(z=0.65) = 0.6817$ \\
$s_{\text{B}, 5}$ & $s_{\text{B}}(z=0.79) = 0.7839$ \\
$s_{\text{B}, 6}$ & $s_{\text{B}}(z=0.95) = 0.8974$\\
$s_{\text{B}, 7}$ & $s_{\text{B}}(z=1.13) = 1.0224$ \\
$s_{\text{F}, 1}$ & $s_{\text{F}}(z=0.25) = -0.1618$ \\
$s_{\text{F}, 2}$ & $s_{\text{F}}(z=0.38) = -0.1279$ \\
$s_{\text{F}, 3}$ & $s_{\text{F}}(z=0.51) = -0.1269$ \\
$s_{\text{F}, 4}$ & $s_{\text{F}}(z=0.65) = -0.1209$ \\
$s_{\text{F}, 5}$ & $s_{\text{F}}(z=0.79) = -0.1164$ \\
$s_{\text{F}, 6}$ & $s_{\text{F}}(z=0.95) = -0.1120$\\
$s_{\text{F}, 7}$ & $s_{\text{F}}(z=1.13) = -0.1080$ \\
\end{tabular}
\end{table}

\begin{table}[H]
\centering
\caption{\itt{List of free parameters considered with their fiducial values in the $\Lambda$CDM model. All the parameters were varied in the Fisher analysis.}}
\label{tab:fids}
\begin{tabular}
{C{1.8cm}|C{6.5cm}}
\hline
Parameter  & Fiducial value \\ 
\hline
$\hat{J}_1$ & $\hat{J}(z=0.25)=\Omega_{\rm m}\cdot\sigma_8(z=0.25) = 0.3388$\\ 
$\hat{J}_2$ & $\hat{J}(z=0.38)=\Omega_{\rm m}\cdot\sigma_8(z=0.38) =0.3666$\\ 
$\hat{J}_3$ & $\hat{J}(z=0.51)=\Omega_{\rm m}\cdot\sigma_8(z=0.51) =0.3846$\\ 
$\hat{J}_4$ & $\hat{J}(z=0.65)=\Omega_{\rm m}\cdot\sigma_8(z=0.65) =0.3949$\\ 
$\hat{J}_5$ & $\hat{J}(z=0.79)=\Omega_{\rm m}\cdot\sigma_8(z=0.79) =0.3979$\\ 
$\hat{J}_6$ & $\hat{J}(z=0.95)=\Omega_{\rm m}\cdot\sigma_8(z=0.95) =0.3948$\\ 
$\hat{J}_7$ & $\hat{J}(z=1.13)=\Omega_{\rm m}\cdot\sigma_8(z=1.13) =0.3859$\\ 
$\hat{J}_8$ & $\hat{J}(z=1.35)=\Omega_{\rm m}\cdot\sigma_8(z=1.35) =0.3707$\\ 
$\hat{J}_9$ & $\hat{J}(z=1.7)=\Omega_{\rm m}\cdot\sigma_8(z=1.7) =0.3427$\\ 
$\hat{J}_{10}$ & $\hat{J}(z=2.1)=\Omega_{\rm m}\cdot\sigma_8(z=2.1) =0.3111$\\ 
$\hat{b}_1$ & $\hat{b}(z=0.25)=b\cdot\sigma_8(z=0.25) =1.4462$\\ 
$\hat{b}_2$ & $\hat{b}(z=0.38)=b\cdot\sigma_8(z=0.38) =1.3512$\\ 
$\hat{b}_3$ & $\hat{b}(z=0.51)=b\cdot\sigma_8(z=0.51) =1.2644$\\ 
$\hat{b}_4$ & $\hat{b}(z=0.65)=b\cdot\sigma_8(z=0.65) =1.1796$\\ 
$\hat{b}_5$ & $\hat{b}(z=0.79)=b\cdot\sigma_8(z=0.79) =1.1035$\\ 
$\hat{b}_6$ & $\hat{b}(z=0.95)=b\cdot\sigma_8(z=0.95) =1.0259$\\ 
$\hat{b}_7$ & $\hat{b}(z=1.13)=b\cdot\sigma_8(z=1.13) =0.9492$\\ 
$\hat{b}_8$ & $\hat{b}(z=1.35)=b\cdot\sigma_8(z=1.35) =0.8683$\\ 
$\hat{b}_9$ & $\hat{b}(z=1.7)=b\cdot\sigma_8(z=1.7) =0.7631$\\ 
$\hat{b}_{10}$ & $\hat{b}(z=2.1)=b\cdot\sigma_8(z=2.1) =0.6690$\\ 
$A_{\rm IA}$ & 1.0\\ 
$\hat{I}_1$ & $\hat{I}(z=0.25)=\Omega_{\rm m}\cdot\sigma_8(z=0.25) = 0.3388$\\ 
$\hat{I}_2$ & $\hat{I}(z=0.38)=\Omega_{\rm m}\cdot\sigma_8(z=0.38) =0.3666$\\ 
$\hat{I}_3$ & $\hat{I}(z=0.51)=\Omega_{\rm m}\cdot\sigma_8(z=0.51) =0.3846$\\ 
$\hat{I}_4$ & $\hat{I}(z=0.65)=\Omega_{\rm m}\cdot\sigma_8(z=0.65) =0.3949$\\ 
$\hat{I}_5$ & $\hat{I}(z=0.79)=\Omega_{\rm m}\cdot\sigma_8(z=0.79) =0.3979$\\ 
$\hat{I}_6$ & $\hat{I}(z=0.95)=\Omega_{\rm m}\cdot\sigma_8(z=0.95) =0.3948$\\ 
$\hat{I}_7$ & $\hat{I}(z=1.13)=\Omega_{\rm m}\cdot\sigma_8(z=1.13) =0.3859$\\ 
$\hat{f}_1$ & $\hat{f}(z=0.25)=f\cdot\sigma_8(z=0.25) = 0.4761$\\
$\hat{f}_2$ & $\hat{f}(z=0.38)=f\cdot\sigma_8(z=0.38) = 0.4826$\\
$\hat{f}_3$ & $\hat{f}(z=0.51)=f\cdot\sigma_8(z=0.51) = 0.4811$\\
$\hat{f}_4$ & $\hat{f}(z=0.65)=f\cdot\sigma_8(z=0.65) = 0.4733$\\
$\hat{f}_5$ & $\hat{f}(z=0.79)=f\cdot\sigma_8(z=0.79) = 0.4612$\\
$\hat{f}_6$ & $\hat{f}(z=0.95)=f\cdot\sigma_8(z=0.95) = 0.4444$\\
$\hat{f}_7$ & $\hat{f}(z=1.13)=f\cdot\sigma_8(z=1.13) = 0.4238$\\
$\hat{b}_{\text{B},1}$ & $\hat{b}_{\text{B}}(z=0.25)=b_{\text{B}}\cdot\sigma_8(z=0.25) = 0.8486$\\
$\hat{b}_{\text{B},2}$ & $\hat{b}_{\text{B}}(z=0.38)=b_{\text{B}}\cdot\sigma_8(z=0.38) = 0.8416$\\
$\hat{b}_{\text{B},3}$ & $\hat{b}_{\text{B}}(z=0.51)=b_{\text{B}}\cdot\sigma_8(z=0.51) = 0.8380$\\
$\hat{b}_{\text{B},4}$ & $\hat{b}_{\text{B}}(z=0.65)=b_{\text{B}}\cdot\sigma_8(z=0.65) = 0.8382$\\
$\hat{b}_{\text{B},5}$ & $\hat{b}_{\text{B}}(z=0.79)=b_{\text{B}}\cdot\sigma_8(z=0.79) = 0.8430$\\
$\hat{b}_{\text{B},6}$ & $\hat{b}_{\text{B}}(z=0.95)=b_{\text{B}}\cdot\sigma_8(z=0.95) = 0.8541$\\
$\hat{b}_{\text{B},7}$ & $\hat{b}_{\text{B}}(z=1.13)=b_{\text{B}}\cdot\sigma_8(z=1.13) = 0.8739$\\
$\hat{b}_{\text{F},1}$ & $\hat{b}_{\text{F}}(z=0.25)=b_{\text{F}}\cdot\sigma_8(z=0.25) = 0.1256$\\
$\hat{b}_{\text{F},2}$ & $\hat{b}_{\text{F}}(z=0.38)=b_{\text{F}}\cdot\sigma_8(z=0.38) = 0.1661$\\
$\hat{b}_{\text{F},3}$ & $\hat{b}_{\text{F}}(z=0.51)=b_{\text{F}}\cdot\sigma_8(z=0.51) = 0.2059$\\
$\hat{b}_{\text{F},4}$ & $\hat{b}_{\text{F}}(z=0.65)=b_{\text{F}}\cdot\sigma_8(z=0.65) = 0.2486$\\
$\hat{b}_{\text{F},5}$ & $\hat{b}_{\text{F}}(z=0.79)=b_{\text{F}}\cdot\sigma_8(z=0.79) = 0.2914$\\
$\hat{b}_{\text{F},6}$ & $\hat{b}_{\text{F}}(z=0.95)=b_{\text{F}}\cdot\sigma_8(z=0.95) = 0.3413$\\
$\hat{b}_{\text{F},7}$ & $\hat{b}_{\text{F}}(z=1.13)=b_{\text{F}}\cdot\sigma_8(z=1.13) = 0.3995$\\
\end{tabular}
\end{table}


\end{document}